\documentclass[aps,prd,showpacs,nofootinbib,amsmath,superscriptaddress]{revtex4}
\usepackage{bm}
\usepackage{graphicx}

\begin{document}

\title{On the importance of Lorentz structure in the parton model:\\
 target mass corrections, transverse momentum dependence, positivity bounds}

\author{U.~D'Alesio}
 \email{umberto.dalesio@ca.infn.it}
 \affiliation{Dipartimento di Fisica, Universit\`a di Cagliari, Cittadella Universitaria,
 I-09042 Monserrato (CA), Italy}
 \affiliation{Istituto Nazionale di Fisica Nucleare, Sezione di Cagliari, C.P. 170,
 I-09042 Monserrato(CA), Italy}

\author{Elliot Leader}
 \email{e.leader@imperial.ac.uk}
 \affiliation{High Energy Physics, Imperial College London, London SW7 2AZ, UK}
 \affiliation{Dipartimento di Fisica, Universit\`a di Cagliari, Cittadella Universitaria,
 I-09042 Monserrato (CA), Italy}

\author{F.~Murgia}
 \email{francesco.murgia@ca.infn.it}
 \affiliation{Istituto Nazionale di Fisica Nucleare, Sezione di Cagliari, C.P. 170,
 I-09042 Monserrato(CA), Italy}

\date{\today}

\begin{abstract}
We show  that respecting the underlying Lorentz structure in the parton
model has very strong consequences. Failure to insist on the correct Lorentz
covariance is responsible for the existence of contradictory results
in the literature for the polarized structure function $g_2(x)$,
whereas with the correct imposition  we are able to derive the
Wandzura-Wilczek relation for $g_2(x)$ and the target-mass
corrections for polarized deep inelastic scattering \emph{without}
recourse to the operator product expansion.
We comment briefly on the problem of threshold behaviour in the presence of target-mass
corrections.
Careful attention to the
Lorentz structure has also profound implications for the structure of the
transverse momentum dependent parton densities often used in parton
model treatments of hadron production, allowing the $\bm{k}_T$
dependence to be \textit{derived explicitly}. It also leads to
stronger positivity and Soffer-type bounds than usually utilized for
the collinear densities.
\end{abstract}

\pacs{11.55.Hx, 11.80.Cr, 12.38.-t, 13.60.Hb, 13.88.+e, 14.20.Dh }

\maketitle

\section{\label{intro} Introduction}
In this paper we shall show that imposition of the correct Lorentz
structure in the parton model allows us to relate
two issues which, \textit{a priori}, do not seem to be connected
with each other: the derivation of a consistent expression for the
polarized deep inelastic scattering (DIS) structure function
$g_2(x)$ in the parton model, and the derivation of the higher twist
target-mass corrections i.e.~corrections of the form $M^2/Q^2$,
where $M$ is the nucleon mass, to $g_1(x,Q^2)$ and $g_2(x,Q^2)$.

As will be discussed below, the target-mass corrections have
previously been derived in a very complicated way from the operator
product expansion (OPE), and the derivation of the Wandzura-Wilczek~(WW)
expression for $g_2(x)$ has involved a questionable analytic
continuation in the OPE moments~\cite{Wandzura:1977qf}. It is thus particularly interesting
that these results can be derived in a field theoretic context
without use of the OPE. Target-mass corrections for
\textit{unpolarized} DIS were first derived by Nachtmann
\cite{Nachtmann:1973mr} employing a very elegant mathematical
approach in which the power series expansion used in the OPE was
replaced by an expansion into a series of hyperspherical functions
(representation functions of the homogeneous Lorentz group). Later,
also within the context of the OPE, Georgi and Politzer
\cite{Georgi:1976ve} re-derived Nachtmann's results using what they
called an alternative analysis ``for simple-minded souls like
ourselves" i.e.~based on a straightforward power series expansion
but, in fact, requiring a very clever handling of the combinatoric
aspects of the problem.

The derivation of target-mass corrections for \textit{polarized} DIS
turned out to be much more difficult. Several papers
\cite{Matsuda:1979ad,Wandzura:1977ce} succeeded in expressing the
reduced matrix elements $a_n$, $d_n$ of the relevant operators in
terms of combinations of moments of the structure functions, but did
not manage to derive closed expressions for the structure functions
$g_{1,2}$ themselves. The latter was finally achieved in 1997 by
Piccione and Ridolfi \cite{Piccione:1997zh} and later generalized to
weak interaction, charged current reactions, by Bl\"{u}mlein and
Tkabladze \cite{Blumlein:1998nv}. These calculations, based on the
OPE, are extremely complicated, and we shall see presently how the
same results can be obtained in a much simpler field-theoretic
approach.

\begin{figure}
 \includegraphics[width=0.3\textwidth]{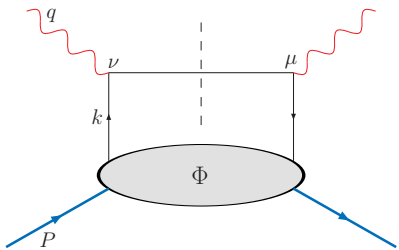}
 \caption{\label{Hand}The DIS ``handbag" diagram involving the $qq$-correlator.}
 \end{figure}

\begin{figure}
 \includegraphics[width=0.4\textwidth]{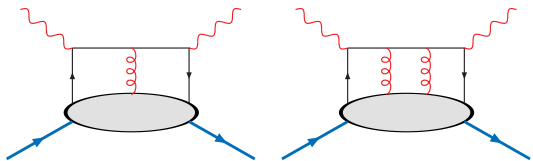}
 \caption{\label{qqG}DIS Diagrams involving the $qqG$ and $qqGG$-correlators.}
 \end{figure}

The clue to this entire approach is contained in the classic paper
of Ellis, Furmanski and Petronzio (EFP) \cite{Ellis:1982cd}, which
gave the first derivation of the dynamic higher twist corrections to
unpolarized DIS in terms of amplitudes involving not just the
``handbag" diagram of Fig.~\ref{Hand}, whose soft part is the
quark-quark $(qq)$ correlator $\Phi $, but the higher order diagrams
in Fig.~\ref{qqG}, whose soft parts are the $qqG$ and $qqGG$
correlators respectively.

EFP begin with a brief discussion of a parton model, which they
refer to as a ``reference model", in which the active quark,
momentum $k^\mu$, emitted from the nucleon is on mass-shell.
Handling the kinematics exactly they arrive at expressions for the
unpolarized structure functions $F_{ 1,2}$ in terms of the quark
densities $q(x)$ which are identical to those of Nachtmann
\cite{Nachtmann:1973mr}.

At first sight it seems surprising that such a naive model should
give the exact results of the theory. But the point is -- and this
is something not stressed in the literature -- that any
off-shellness of the quark is a direct consequence of QCD, i.e.~if
the strong interaction coupling $g=0$ then it follows that
(taking the quark mass $m_q=0$) $k^2=0$
[see Eq.~(\ref{E})]. Target-mass corrections are, by definition,
kinematic in origin, therefore are independent of the value of $g$.
Hence it is not miraculous that a model with $k^2=0$,
i.e.~equivalent to putting $g=0$, should yield exact results for the
target-mass corrections. However, it is crucial that the Lorentz
structure is built into the model, as is done by EFP
\cite{Ellis:1982cd}. The implications of this are that the exact
target-mass corrections must be derivable from the ``handbag"
diagram, Fig.~\ref{Hand}, alone, since the diagrams of
Fig.~\ref{qqG} vanish when $g=0$. \newline We shall carry out the
derivation for polarized DIS and also show that when the Lorentz
structure is respected the $k^2=0$ ``model" yields an unambiguous
result for $g_2(x)$, namely the Wandzura-Wilczek (WW) result
\cite{Wandzura:1977qf}
\begin{equation}\label{A}
g_2(x_{\rm Bj}) = -g_1(x_{\rm Bj}) + \int_{x_{\rm Bj}}^1 {\rm d}y\>\frac{g_1(y)}{y} \>,
\end{equation}
where $x_{\rm Bj}$ is the well-known Bjorken variable for DIS.
This suggests that the analytic continuation necessary in the OPE
derivation of the WW result is in fact correct, and also implies
that in a correctly formulated parton model, with on-shell quarks,
$g_2(x)$ is exactly given by the WW expression.

In this paper we shall also show that careful attention to the Lorentz
structure in the parton model imposes strong constraints on the
possible $\bm{k}_T$ dependence of the so-called transverse momentum dependent
(TMD) parton densities, that in recent years have received
much emphasis (for an up-to-date review see e.g.~Ref.~\cite{D'Alesio:2007jt}
and references therein). Indeed, taking into account the
transverse momentum of partons is important for an understanding of
the large transverse single spin asymmetries observed in many
reactions. It is also essential in order to generate the parton
orbital angular momentum which appears necessary as a consequence of
the small contribution to the nucleon angular momentum provided by
the parton spins.

As we are going to show, imposing Lorentz covariance we are able to \textit{derive},
for the unpolarized densities
$q(x,\,\bm{k}^2_T)$, the longitudinal densities $\Delta
q(x,\,\bm{k}^2_T)$ and the transversity densities $\Delta _T
q(x,\,\bm{k}_T)$ or $\Delta '_T q(x,\,\bm{k}^2_T)\equiv
h_1(x,\,\bm{k}^2_T)$, their  dependence
on $\bm{k}_T$ from the functional form of the usual collinear,
$\bm{k}_T$-integrated, parton densities $ q(x)$, $\Delta q(x)$ and
$\Delta _T q(x)$ [These relations are spelled out in detail in
Eqs.~(\ref{VAA}),~(\ref{WA2}) and (\ref{WA}), and in
Eqs.~(\ref{GB}),~(\ref{Q2})]. Moreover, if  perturbative evolution
with $Q^2$ is taken into account, then the evolution of the
$\bm{k}_T$-dependent densities is entirely controlled via the known
evolution of the collinear densities.

It should be noted that the functional form found in the parton
model is quite different from the often used factorized form,
typically  $f(x)\,e^{-\lambda \, \bm{k}_T ^2}$, though the latter may be a reasonable
starting point for analyzing the presently available data.

Further, we obtain positivity and Soffer-type
bounds~\cite{Soffer:1994ww}, based on the $\bm{k}_T$-dependent
densities, which are stronger, in general, than those usually
imposed on the purely $x$-dependent, $\bm{k}_T$-integrated, collinear densities, and we
suggest that it would be interesting to impose these bounds on the collinear densities when
extracting parton densities from deep inelastic and semi-inclusive
deep inelastic scattering data.

Therefore, as will become even more clear in the following, careful attention to the
Lorentz structure in the parton model with on-shell partons,
i.e.~with $k^2=0$, has dramatic consequences. Given the standard
belief that the \textit{soft} functions appearing in QCD have
support only in a very narrow range of values of $k^2$ around
$k^2=0$, it is tempting to suppose that the results derived in this
paper may not be too different from those holding in a full QCD
treatment.

Covariant parton models have been discussed in the literature for
many decades, beginning with the work of Landshoff, Polkinghorne and Short
in 1971~\cite{Landshoff:1970ff}, and Franklin in 1977~\cite{Franklin:1976dm}.
These papers dealt only with the unpolarized structure functions $F_{1,2}(x)$.
The collinear spin-dependent structure functions $g_{1,2}(x)$  were studied,
in response to the ``spin crisis in the parton model", by Jackson, Roberts and
Ross~\cite{Jackson:1989ph}, who also commented upon some aspects of quark
transverse momentum. More recently, a very detailed study of covariant models
was initiated by Zavada in 1997~\cite{Zavada:1996kp}, and later developed by
Zavada~\cite{Zavada:2002uz,Zavada:2007ww} and Efremov, Teryaev and Zavada~\cite{Efremov:2004tz},
to include the polarized parton densities as well as parton transverse momentum.
While our paper was in its last phase of preparation we became aware of a
further paper by Efremov, Schweitzer, Teryaev and Zavada~\cite{Efremov:2009ze}
(which we shall refer to as ESTZ) and a very latest work by Zavada~\cite{Zavada:2009sk}
in which TMD parton distributions are discussed in detail.

In these papers the analysis is based upon reasonable assumptions about the structure
of the quark wave function in the nucleon \emph{rest} frame. Because the treatment
is covariant the results can then be translated to any other Lorentz frame.
It is remarkable that many of the results of this kind of analysis are in complete
agreement with our more general, frame independent, treatment. The principal differences
are that: \textit{i}) in our treatment the three leading twist densities, the unpolarized density
$q(x,\,\bm{k}^2_T)$, the longitudinal density $\Delta q(x,\,\bm{k}_T^2)$ and the
transversity density $\Delta _T q(x,\,\bm{k}_T)$ or
$\Delta '_T q(x,\,\bm{k}^2_T)\equiv h_1(x,\,\bm{k}^2_T)$ are \emph{independent},
whereas in the Zavada et al. papers only two of these are independent;
\textit{ii}) we take into account target mass corrections.
Because our approaches are so different, we believe it will be instructive
to present our derivation in full.

The plan of the paper is the following: in section~\ref{corr} we
discuss the hadronic correlator for polarized DIS in the parton
model, with $g=0$, and give the general expression of the structure
functions $g_1(x)$ and $g_2(x)$; in section~\ref{coeff} we evaluate
explicitly $g_1(x)$ and $g_2(x)$ including target-mass corrections and
derive the WW relation; in section~\ref{tmd} we extend our analysis
to the TMD quark distributions and discuss the consequences of
imposing Lorentz invariance; in section~\ref{bounds} we examine the
positivity and Soffer bounds for the $\bm{k}_T$-dependent
unpolarized, longitudinally and transversely polarized quark
distributions, discuss how their correct implementation leads to
new, more stringent, bounds on the $\bm{k}_T$-integrated collinear
distributions and show some phenomenological implications of
this approach; finally, in section~\ref{concl} we give some additional comments and
conclusions.

\section{\label{corr} The hadronic correlator $\Phi$ for $g=0$}

The correlator $\Phi_{ij}(P,S;k)$, where $P^\mu$ and $S^\mu$ are the
momentum and spin-polarization 4-vectors for the nucleon ($S^2 =
-1$) , $k^\mu$ is the quark 4-momentum and $i,j$ are Dirac indices,
is defined by
\begin{equation} \label{B}
\Phi_{ij}(P,S;k) = \int\,\frac{{\rm d}^4z}{(2\pi)^4}\, e^{ik\cdot z}\,
 \langle P,S|\bar{\psi}_j(0) \, \psi_i(z)|P,S\rangle\>.
\end{equation}
To preserve colour gauge invariance a path-ordered Wilson line (or
gauge-link) should be inserted between the quark fields. Since we
are mostly interested in the $g=0$ case, we will take this operator
as the identity.

We ignore for the moment flavour and the quark charge -- they are
trivially reinstated at the end -- and, as already stated, work with $m_q=0$. By partial
integration we see that, when $g=0$,
\begin{equation}
{\rm Tr}[\Phi {\not\! k}] \>  =\>  i \,\int\,\frac{{\rm
d}^4z}{(2\pi)^4}\, e^{ik\cdot z}\, \langle P,S|\bar{\psi}(0)
{\not\!\partial } \, \psi(z)|P,S\rangle \>=\> 0\>, \label{C}
\end{equation}
and that for any of the 16 Dirac matrices $\Gamma $ ($=
I$, $\gamma^\mu$, $\sigma^{\mu\nu}$, $\gamma^\mu\gamma_5$,
$i\gamma_5$),
\begin{equation} \label{D}
{\rm Tr}[\Phi \, \Gamma \,{\not\! k}] = 0\>.
\end{equation}

Finally, again by partial integration, we get for all $k^2$
\begin{equation} \label{E}
k^2 \, \Phi_{ij} =  - \,\int\,\frac{{\rm d}^4z}{(2\pi)^4}\,
e^{ik\cdot z}\, \langle P,S|\bar{\psi}_j(0) {\not\!\partial }^2 \,
\psi_i(z)|P,S\rangle = 0\>,
\end{equation}
implying that
\begin{equation} \label{F}
\Phi \propto \delta (k^2)\>.
\end{equation}
$\Phi $ is a $4\times 4$ matrix in Dirac spin space. The first
attempt to write down its most general form was made by Ralston and
Soper \cite{Ralston:1979ys} who, however, missed one term
\cite{Leader:1979ys}. The full dynamical twist-two structure, including the three
so-called ``naively $T$-odd" amplitudes, contains twelve terms~\cite{Mulders:1995dh}.
Eight of these, including all the
``naively $T$-odd" amplitudes, are eliminated by the requirements of
Eqs.~({\ref{C})-(\ref{E}}) and two become related to each other. We
are left with
\begin{equation} 
\lefteqn \, \Phi \,(P, S;k) \stackrel{g=0}{=} A_3\, {\not\! k} +
\frac{A_8}{M}\, (k\cdot S)\, {\not\! k}\, \gamma_5 +
\frac{A_{11}}{M^2}\, i \,\sigma ^{\mu\nu} \,\gamma _5 \,k_\mu
\,[(k\cdot S)\,P_\nu - (k\cdot P)\,S_\nu ]\>. \label{H}
\end{equation}
The scalar functions $A_{3,8,11}$ are, in general, functions of
$P\cdot k$ and $k^2$. In our case, bearing in mind Eq.~(\ref{F}), we
shall later use
\begin{equation} \label{O}
A_3 = \frac{1}{\pi M^2}\, \varphi _3 \big(\frac{2P\cdot k}{M^2}\big)\, \delta (k^2)\,\theta [(P-k)^2]
\end{equation}
and
\begin{equation} \label{P}
A_{8,11} = -\,\frac{2}{\pi M^2}\, \varphi _{8,11} \big(\frac{2P\cdot k}{M^2}\big)
\, \delta (k^2)\,\theta [(P-k)^2]\>,
\end{equation}
where the factors are for later convenience and, following
Ref.~\cite{Ellis:1982cd}, the $\theta $-function ensures that the
nucleon remnant has positive energy. Notice again the factor
$\delta(k^2)$ in these expressions.

 For the moment we shall only deal with the  part relevant to $g_{1,2}(x_{\rm Bj})$ i.e.~$A_8$.
Polarized DIS is controlled by the antisymmetric part $W_A^{\mu\nu}$
of the hadronic tensor, related to the structure functions
$g_{1,2}(x_{\rm Bj})$ via~\cite{Anselmino:1994gn}
\begin{equation} \label{H2}
W_A^{\mu \nu}  =   \frac{2M}{P\cdot q}\,\epsilon^{\mu \nu \alpha
\beta}\, q_\alpha \,\bigg\{ S_\beta \,g_1
   +  \Big[\,S_\beta - \frac{(S\cdot q)\,P_\beta}{P\cdot q}\,\Big]\, g_2  \bigg\} \>,
\end{equation}
where $q$ is the photon 4-momentum and our convention is $\epsilon_{0123} = 1$.

The contribution from the handbag diagram, Fig.~\ref{Hand}, to $W_A^{\mu\nu}$  is then
\begin{eqnarray} \label{I}
 \! \! \! W_A^{\mu\nu} &= & \epsilon^{\mu \nu \rho \sigma }
 \int {\rm d}^4k \,(k_\rho + q_\rho)\,\delta [(k + q)^2]
 \, {\rm Tr}(\gamma_\sigma\,\gamma_5\,\Phi) \nonumber \\
 &= & -\,\epsilon^{\mu \nu \rho \sigma }\int {\rm d}^4k\,(k_\rho + q_\rho)\,\delta [(k + q)^2]
  \,{\rm Tr}(\gamma_\sigma\,{\not\! k})\, \frac{A_8}{M}\,(k\cdot S) \nonumber \\
 &=& - \,\frac{4}{M}\,\epsilon^{\mu \nu \rho \sigma }q_\rho\,S^\beta I_{\beta \sigma}\>,
\end{eqnarray}
where
\begin{equation} \label{J}
I_{\beta \sigma} = \int {\rm d}^4k \,k_\beta \,k_\sigma \,\delta [(k + q)^2] \,A_8\>.
\end{equation}
Note that the hadronic tensor $W_A^{\mu\nu}$ in Eq.~(\ref{I}) is electromagnetic gauge invariant.

We introduce the standard two auxiliary null vectors $p^\mu$ and
$n^\mu$ , $p^2 = n^2 =0$, $p\cdot n=1$, in a slightly unusual way
via
\begin{equation} \label{K}
p^\mu = \frac{1}{D}\,(P^\mu - \frac{M^2 \xi}{Q^2}\, q^\mu ) \qquad \qquad \qquad
n^\mu = \frac{2 \xi}{D Q^2}\, (q^\mu + \xi \,P^\mu)\>,
\end{equation}
where $\xi $ is the Nachtmann variable
\begin{equation} \label{M}
\xi = \frac{2 x_{\rm Bj}}{1 + \sqrt{1 + 4M^2 x_{\rm Bj}^2/Q^2}}\>,
\end{equation}
and
\begin{equation} \label{N}
D \equiv 1 + \epsilon  \qquad \qquad \qquad \epsilon \equiv \frac{M^2 \xi ^2}{Q^2}\>.
\end{equation}
Note that
\begin{equation} \label{N1}
q\cdot p = \frac{Q^2}{2\xi}   \qquad \qquad q\cdot n = -\xi
\qquad \qquad
x_{\rm Bj} = \frac{\xi}{N} \qquad \qquad N\equiv 1-\epsilon \>.
\end{equation}
Switching off the target-mass corrections thus corresponds to
putting $\epsilon = 0, \,D = N = 1$.

Since by definition, see Eq.~(\ref{J}), $I_{\beta\sigma}$ is
symmetric under $\beta\longleftrightarrow\sigma$, its most general
expression can be written as
\begin{equation}
I_{\beta\sigma} = M^2B_{g}\,g_{\beta\sigma}+M^2B_{pn}\,(\,p_\beta n_\sigma+p_\sigma n_\beta\,)
+B_{p}\,p_\beta p_\sigma + M^4B_{n}\, n_\beta n_\sigma\>.
\label{Igen}
\end{equation}
On the other hand, from Eq.~(\ref{J}), bearing in mind the factor
$\delta(k^2)$ in $A_8$, Eq.~(\ref{P}), and introducing the shorthand
\begin{equation} \label{U}
\langle\, X\, \rangle \equiv \int {\rm d}^4k \,X \,\delta[(k + q)^2] \,A_8\>,
\end{equation}
one finds that
\begin{eqnarray}
g^{\beta\sigma}\,I_{\beta\sigma}  &=& \langle\,k^2\,\rangle = 0 \nonumber\\
p^\beta n^\sigma\,I_{\beta\sigma} &=& \langle\,(k\cdot p)\,(k\cdot n)\,\rangle \>\equiv\> M^2C_{pn} \nonumber \\
p^\beta p^\sigma\,I_{\beta\sigma} &=& \langle\,(k\cdot p)^2\,\rangle \>\equiv\> M^4C_{p}\label{c-coeff}\\
n^\beta n^\sigma\,I_{\beta\sigma} &=& \langle\,(k\cdot n)^2\,\rangle \>\equiv\> C_{n} \>. \nonumber
\end{eqnarray}
By comparing with Eq.~(\ref{Igen}) and expressing the factors
$B_{i}$ in terms of the $C_{i}$'s one can then write
\begin{equation} \label{Q}
I_{\beta \sigma} = M^2C_{pn}\,\big[\,2(p_\beta n_\sigma + p_\sigma n_\beta)-g_{\beta\sigma}\,\big] + C_{n}\, p_\beta p_\sigma
+ M^4 C_{p}\, n_\beta n_\sigma\>.
\end{equation}

Moreover, from Eq.~(\ref{K}), because of the factor $\epsilon^{\mu \nu \rho \sigma
}q_\rho\,S^\beta $ in Eq.~(\ref{I}) and the fact that $ P\cdot
S=0$, we can replace\newline
\begin{center}
$p_\sigma \rightarrow \frac{1}{D}P_\sigma \qquad n_\sigma
\rightarrow \frac{2\xi^2}{D Q^2}P_\sigma \qquad p_\beta \rightarrow
-\frac{M^2 \xi}{D Q^2}q_\beta  \qquad  n_\beta \rightarrow
\frac{2\xi}{D Q^2}q_\beta $
\end{center}
in Eq.~(\ref{Q}). Thus
\begin{equation} \label{R}
I_{\beta \sigma} \rightarrow -M^2\, C_{pn} \, g_{\beta\sigma} + \frac{\xi M^2}{D^2 Q^2}
\big[\,4N C_{pn} -C_{n} +4\,\epsilon\, C_{p}\,\big]\, q_\beta P_\sigma\>.
\end{equation}
Putting this into Eq.~(\ref{I}) and comparing with Eq.~(\ref{H2}),
we can read off directly expressions for $g_1 + g_2$ and $g_2$ and
thereby obtain
\begin{equation} \label{S}
g_1(x_{\rm Bj}) + g_2(x_{\rm Bj}) = 2\,(P\cdot q)\, C_{pn}\>,
\end{equation}
and
\begin{equation} \label{T}
g_1(x_{\rm Bj}) = 2 \,(P\cdot q) \,\Big\{\,C_{pn} - \frac{N}{2D^2}\,\big[\,4N\,C_{pn} - C_{n} + 4\,\epsilon \,
 C_{p}\,\big]\,\Big\}\>,
\end{equation}
where we have used Eq.~(\ref{N1}).
\section{\label{coeff} Target-mass corrections and the Wandzura-Wilczek relation}
In order to evaluate the coefficients $C_{pn}$, $C_{p}$, $C_{n}$ in Eqs.~(\ref{S}),~(\ref{T}) we
parameterize $k^\mu$ in the usual way, bearing in mind that $k^2=0$,
\begin{equation} \label{W}
k^\mu = x\, p^\mu + \frac{\bm {k}^2_T}{2x}\,n^\mu + k_T^\mu\>,
\end{equation}
where $k_T^2 = - \bm{k}^2_T$.
Then
\begin{equation} \label{X}
\int \, {\rm d}^4k = \frac{1}{2}\,\int \frac{{\rm d}x}{x}\, {\rm d}^2 \bm{k}_T \,{\rm d}k^2 =
 \frac{\pi}{2} \int \frac{{\rm d}x}{x}\, {\rm d}\bm{k}_T^2 \,{\rm d}k^2\>,
\end{equation}
and in terms of these variables
\begin{equation} \label{Y}
C_{pn}=\langle\,\frac{\bm{k}_T^2}{2M^2}\> \rangle \qquad C_{n} = \langle\> x^2\, \rangle
\qquad C_{p} =\langle\, \Bigl[\frac{\bm{k}_T^2}{2xM^2}\Bigr]^2\, \rangle\>.
\end{equation}
Bearing in mind Eqs.~(\ref{N1}) and (\ref{W}) one has
\begin{equation} \label{Z}
\delta [(k + q)^2] = \frac{x}{\xi}\, \delta \big[ \bm{k}_T^2 - Q^2
\,\frac{x}{\xi}(\frac{x}{\xi} - 1)\big]\>.
\end{equation}
We therefore integrate over $ \bm{k}_T^2 $, introduce
\begin{equation} \label{AA}
\eta \equiv \frac{2P\cdot k}{M^2}
= x + \frac{\bm{k}_T^2}{xM^2}
= \frac{Q^2}{M^2 \xi^2}\,[\, D \,x - \xi\, ]\>,
\end{equation}
via Eq.~(\ref{Z}), and have
\begin{equation} \label{BA}
\int {\rm d}x = \frac{\epsilon}{D}\, \int _\xi^1 \, {\rm d}\eta\>,
\end{equation}
where the upper integration limit comes from the $\theta
$-function in Eqs.~(\ref{O}), (\ref{P}), and the lower limit from $\eta
\geq x \geq \xi $. In terms of $\eta $ the variables appearing in
Eq.~(\ref{Y}) are given by
\begin{equation} \label{CA}
x=\frac{1}{D}(\xi + \epsilon\eta) \qquad  \qquad
\frac{\bm{k}_T^2}{xM^2} = \frac{1}{D}(\eta - \xi)\>.
 \end{equation}
We substitute the expressions for the coefficients, Eq.~(\ref{Y}),
into Eqs.~(\ref{S}), (\ref{T}) using Eq.~(\ref{P}), restore the
quark charge, separate the leading twist terms, i.e.~those that do
not vanish when $ \epsilon \rightarrow 0 $, and, inside the integral, order the rest
according to the powers of $\epsilon$ and $\eta $ involved, to
obtain, for each flavour,

\begin{equation} \label{DA}
g_1(x_{\rm Bj})= e^2_q \,\frac{N}{2D^5}\, \int _\xi^1 \, {\rm d}\eta \,
 \Big\{\, \xi(-2\xi+\eta) + \epsilon\,(8\xi^2-11\xi\eta+2\eta^2)
 +\epsilon^2(-2\xi^2+11\xi\eta-8\eta^2)+\epsilon^3(-\xi\eta+2\eta^2) \Big\}\>\varphi_8(\eta)\>,
 \end{equation}

\begin{equation} \label{EA}
 g_1(x_{\rm Bj}) + g_2(x_{\rm Bj})= e^2_q \,\frac{N}{2D^3}\, \int_\xi^1 \,
 {\rm d}\eta \,\Big\{\, \xi\,(\xi - \eta)
  +\epsilon \, \eta\,(\xi - \eta)\Big\}\> \varphi_8(\eta)\>.
\end{equation}

{}From here we also get a separate parton-model expression for $g_2$:
\begin{equation} \label{G2}
 g_2(x_{\rm Bj}) = e^2_q \,\frac{N}{2D^5}\, \int _\xi^1 \, {\rm d}\eta \,\Big\{\, \xi\,(3\xi - 2\eta)
  +\epsilon \,(-6\xi^2 + 10\xi\eta - 3\eta^2)+\epsilon^2(3\xi^2-10\xi\eta+6\eta^2)
  +\epsilon^3(2\xi\eta-3\eta^2)\Big\}\> \varphi_8(\eta)\>.
\end{equation}
 \subsection{\label{coeff-nomass} Neglecting target-mass corrections}
 The expression for $g_1(x_{\rm Bj})$, Eq.~(\ref{DA}), must  agree
with the standard collinear parton model result, with no target-mass
corrections~(NTM), in terms of the (longitudinal) polarized quark
density,
 \begin{equation} \label{FA}
 g_1^{\rm NTM}(x_{\rm Bj}) = \frac{e^2_q}{2} \, \Delta q(x_{\rm Bj})\>,
 \end{equation}
when target masses are neglected, i.e.~when $\epsilon \rightarrow 0$ and $\xi \rightarrow x_{\rm Bj} $.
Thus
\begin{equation} \label{GA}
\Delta q(x_{\rm Bj}) = x_{\rm Bj} \,\int _{x_{\rm Bj}}^1 \, {\rm
d}\eta \,(\eta - 2\,x_{\rm Bj})\,\varphi_8(\eta)\>.
 \end{equation}

This relation between $ \varphi_8(\eta) $ and $\Delta q(x_{\rm Bj})$
allows us to express the integrals in Eq.~(\ref {DA}) in terms of
integrals over  $\Delta q(\xi)$, but also has powerful consequences
for the structure of transverse momentum dependent parton densities,
as will be discussed shortly.

 Turning to $g_1(x_{\rm Bj}) + g_2(x_{\rm Bj})$, we have from
Eq.~(\ref{EA}):
 \begin{equation} \label{HA}
 g_1^{\rm NTM}(x_{\rm Bj}) + g_2^{\rm NTM}(x_{\rm Bj}) =
\frac{e^2_q}{2}\,x_{\rm Bj}\,\int _{x_{\rm Bj}}^1 \, {\rm d}\eta
\,(x_{\rm Bj}- \eta)\,\varphi_8(\eta)\>.
 \end{equation}
Consider now [see Eq.~(\ref{GA})]
\begin{equation} \label{IA}
 \int _\xi^1 \, \frac{d\eta}{\eta} \,\Delta q(\eta) = \int _\xi^1 \,
d\eta\,\int _\eta^1 \, d\eta'( \eta' - 2\,\eta)\varphi_8(\eta')\>.
  \end{equation}
Changing the order of integration, we obtain ($\xi \leq \eta \leq
\eta' \leq 1$)
  \begin{equation} \label{JA}
 \int _\xi^1 \, \frac{{\rm d}\eta}{\eta} \,\Delta q(\eta) = \xi \,
\int _\xi^1 \, {\rm d}\eta'(\xi - \eta')\,\varphi_8(\eta')\>.
 \end{equation}
Putting this relation in Eq.~(\ref{HA}) and using Eq.~(\ref{FA}) yields
  \begin{equation} \label{KA}
 g_1^{\rm NTM}(x_{\rm Bj}) + g_2^{\rm NTM}(x_{\rm Bj}) = \int_{x_{\rm Bj}}^1
 \frac{{\rm d}x'}{x'}\,g_1^{\rm NTM}(x')\>,
  \end{equation}
which is the Wandzura-Wilczek relation in the absence of target-mass
corrections.
  \subsection{\label{coeff-mass} Inclusion of target-mass corrections}
Returning to Eqs.~(\ref{DA}), (\ref{EA}) we note first that from
Eqs.~(\ref{JA}), (\ref{GA})
\begin{equation} \label{LA}
 \xi\,\int _\xi^1 \, {\rm d}\eta\,\varphi_8(\eta) = -\frac{1}{\xi}\,
 \Big\{\, \Delta q(\xi) +   \int _\xi^1 \, \frac{{\rm d}\eta}{\eta} \,\Delta q(\eta)\, \Big\}\>.
\end{equation}
Then, using again Eq.~(\ref{GA}) in the second step,
\begin{eqnarray} \label{MA}
 \int _\xi^1 \, {\rm d}\eta \,\eta \,\varphi_8(\eta)&=&
 \int _\xi^1 \, {\rm d}\eta\, (\,\eta - 2\xi + 2\xi\,)\varphi_8(\eta) \nonumber \\
 &=& \frac{\Delta q(\xi)}{\xi} +2\,\xi\,\int _\xi^1 \,
 {\rm d}\eta\,\varphi_8(\eta)
 \nonumber \\
 &=& -\frac{1}{\xi}\,\Big\{\,\Delta q(\xi) + 2 \int _\xi^1
  \, \frac{{\rm d}\eta}{\eta} \,\Delta q(\eta)\,\Big\}\>.
\end{eqnarray}
Further, integrating by parts,
\begin{eqnarray} \label{NA}
 \int _\xi^1 \, {\rm d}\eta \,\eta^2 \,\varphi_8(\eta)&=&
 \xi\,\int _\xi^1 \, {\rm d}\eta \,\eta \,\varphi_8(\eta)
  +  \int _\xi^1 \, {\rm d}\eta \, \int _\eta^1 \, {\rm d}\eta' \,\eta' \,\varphi_8(\eta') \nonumber \\
 &=& -\Delta q(\xi) -3\,\int _\xi^1 \, \frac{{\rm d}\eta}{\eta} \,\Delta q(\eta)
 - 2\,\int _\xi^1 \, \frac{{\rm d}\eta}{\eta}\int _\eta^1 \, \frac{{\rm d}\eta'}{\eta'} \,\Delta q(\eta')\>.
\end{eqnarray}
Substituting Eqs.~(\ref{LA})-(\ref{NA}) into Eqs.~(\ref{DA}),~(\ref{EA}) we obtain
expressions which, for reasons to be explained in the next subsection,  we shall call
the \emph{bare parton model} (BPM) results
\begin{equation} \label{OA}
 g_1^{\rm BPM}(x_{\rm Bj}) = \frac{e^2_q}{2}\, \Big\{\,\frac{N^2}{D^3}\,\Delta q(\xi) +
\frac{4\epsilon N(1+N)}{D^4}\int _\xi^1 \, \frac{{\rm d}\xi'}{\xi'} \,\Delta q(\xi')
 -\frac{4\epsilon N(N^2-2\epsilon)}{D^5}\int _\xi^1 \, \frac{{\rm d}\xi'}{\xi'}\int
_{\xi'}^1 \, \frac{{\rm d}\xi''}{\xi''} \,\Delta q(\xi'')\,\Big\}
\end{equation}
\begin{equation} \label{PA}
 g_1^{\rm BPM}(x_{\rm Bj}) + g_2^{\rm BPM}(x_{\rm Bj})= \frac{e^2_q}{2}\, \Big\{\,\frac{N}{D^2}\, \int
_\xi^1 \, \frac{{\rm d}\xi'}{\xi'} \,\Delta q(\xi')+ \frac{2 \epsilon N}{D^3}\,\int
_\xi^1 \, \frac{{\rm d}\xi'}{\xi'}\int _{\xi'}^1 \, \frac{{\rm d}\xi''}{\xi''} \,\Delta
q(\xi'')\,\Big\}
\end{equation}
\begin{equation} \label{G2A}
 g_2^{\rm BPM}(x_{\rm Bj}) = \frac{e^2_q}{2}\, \Big\{-\frac{N^2}{D^3}\,\Delta q(\xi) +
\frac{N^2(N-4\epsilon)}{D^4}\int _\xi^1 \, \frac{{\rm d}\xi'}{\xi'} \,\Delta q(\xi')
 +\frac{6\epsilon N^3}{D^5}\int _\xi^1 \, \frac{{\rm d}\xi'}{\xi'}\int
_{\xi'}^1 \, \frac{{\rm d}\xi''}{\xi''} \,\Delta q(\xi'')\,\Big\}\>,
\end{equation}
which, as can be shown after some algebra, agree exactly
with the dynamical twist-two (i.e.~taking $d_n=0$) OPE results of
Ref.~\cite{Piccione:1997zh}.
 Furthermore we find that
 \begin{equation} \label{QA}
  \xi\,\frac{{\rm d}}{{\rm d}\xi}\,[g_1^{\rm BPM}(x_{\rm Bj}) + g_2^{\rm BPM}(x_{\rm Bj})] = -\frac{D}{ N}\>
g_1^{\rm BPM}(x_{\rm Bj})\>,
  \end{equation}
so that, bearing in mind Eqs.~(\ref{N}) and (\ref{N1}),
  \begin{equation} \label{RA}
  g_1^{\rm BPM}(x_{\rm Bj}) + g_2^{\rm BPM}(x_{\rm Bj})= \int _{x_{\rm Bj}}^1
\frac{{\rm d}x'}{x'}\,g_1^{\rm BPM}(x')\>,
  \end{equation}
and the Wandzura-Wilczek relation holds also when target-mass corrections are
included \cite{Piccione:1997zh}.
\subsection{\label{coeff-th} Target mass corrections and threshold behaviour}
There is a long-standing problem about the behaviour of the structure functions,
both unpolarized and polarized, at $x_{\rm Bj}=1$. To see this most simply
consider the first term in the expansion of Eq.~(\ref{OA}) in powers of $M^2/Q^2$,
where we show the kinematic dependence in explicit detail
\begin{equation} \label{g1zeroth}
g_1(x_{\rm Bj}) = \frac{e_q^2}{2}  \Delta q[\,x=\xi(x_{\rm Bj},Q^2)\,]\>. \end{equation}

It is usually stated that one must have $g_1(x_{\rm Bj}=1)=0$, implying the bizarre
result that $\Delta q(x) $ vanishes at $x=\xi(1,Q^2)$ i.e.~at a continuous infinity
of $Q^2$ dependent points. Of course without target mass corrections the vanishing
of $g_1(x_{\rm Bj}=1)$ simply implies that $\Delta q(x=1)=0\,$.

There are no satisfactory prescriptions for avoiding this issue in the literature.
Georgi and Politzer \cite{Georgi:1976ve}
and Piccione and Ridolfi \cite{Piccione:1997zh} argue that higher twist terms must
be taken into account in the region of large $x_{\rm Bj}$, whereas
Accardi and Melnitchouk \cite{Accardi:2008pc} impose
some constraints on the virtuality of the struck quark.

We would like to propose a very simple resolution to this problem. In the standard
parton model treatment there is the underlying assumption that the struck quark and
target fragments materialize into physical hadrons with  probability one. This is
tantamount to the assumption of \emph{equivalent completeness} of partonic and
hadronic states i.e.
\begin{equation}\label{compl}
\sum_{\substack{\textrm{all  parton} \\ \textrm{states}}} |parton\rangle \langle parton|\>\>=
\sum_{\substack{\textrm{all hadron} \\ \textrm{states}}} |hadron\rangle \langle hadron|\>.
\end{equation}
While this might be reasonable for massless hadrons, it cannot be true when hadrons
possess their physical masses. Thus what is needed is a \emph{non-perturbative}
function to express the failure of equivalent completeness. It has to express the
fact that there is zero probability for the final partonic state (the struck parton
and the target fragments) to hadronize if its total energy  is too small to produce
an inelastic hadronic event, and that the probability for hadronization is one when
the energy is large enough. The simplest possibility is $\theta (x_{\rm Th} - x_{\rm Bj})$,
where $x_{\rm Th}$ is the maximum kinematically allowed value of $x_{\rm Bj}$ for
\emph{inelastic} scattering. We note that, strictly speaking, and excluding photon
production which is higher order in the fine structure constant,  this is \emph{not}
$x_{\rm Th}=1$, but
\begin{equation}\label{xTh}
x_{\rm Th} = \frac{Q^2}{Q^2 + \mu (2M + \mu )}\>, \end{equation}
where  $\mu$ is the pion mass.

Thus we propose that in a \emph{physical parton model} (PPM) the BPM target mass
dependent expressions, Eqs.~(\ref{OA}),~(\ref{PA}), for $g_1^{\rm BPM}$ and $g_1^{\rm BPM} +
g_2^{\rm BPM}$ should be modified to
\begin{equation}\label{modifedg1TMC}
g_1^{\rm PPM}(x_{\rm Bj}) = g_1^{\rm BPM}(x_{\rm Bj}) \, \theta (x_{\rm Th} - x_{\rm Bj})
 \end{equation}
 \begin{equation} \label{modifiedg12TMC}
 g_1^{\rm PPM}(x_{\rm Bj}) + g_2^{\rm PPM}(x_{\rm Bj})=  [g_1^{\rm BPM}(x_{\rm Bj}) + g_2^{\rm BPM}(x_{\rm Bj})]\,
\theta (x_{\rm Th} - x_{\rm Bj})\>.
 \end{equation}
  \section{\label{tmd} Results for the transverse momentum dependent quark densities}
 In the standard collinear parton model the transverse momentum $
\bm{k}_T $ is integrated over up to the large scale $Q$. But in the
last few years many polarized reactions have been studied in a more
general framework in which the fundamental parton densities $q(x),\,
\Delta q(x)$ and $\Delta_T q(x)$ are allowed to depend on the \emph{intrinsic}
$\bm{k}_T$ (see e.g.~Refs.~\cite{D'Alesio:2007jt,Anselmino:2005sh} and references therein).
In all these approaches
it is often assumed that the $x$ and $ \bm{k}_T$ dependence can
be factorized. This, as we shall see, is in contradiction with the
Lorentz structure of the amplitudes, \emph{at least in the $g=0$ case}. We
shall comment later on the more general situation.

For unpolarized DIS the analogue of Eq.~(\ref{GA})  is~\cite{Ellis:1982cd}
\begin{equation} \label{SA}
q(x_{\rm Bj}) = x_{\rm Bj} \,\int _{x_{\rm Bj}}^1 \, {\rm d}\eta \,\,\varphi_3(\eta)\>.
\end{equation}
Moreover, on changing integration variables from $\eta$ to
$\bm{k}_T^2$ via Eq.~(\ref{AA}), we see that $q(x,\bm{k}_T ^2)$ for
which
 \begin{equation} \label{TA}
 \int {\rm d}^2\bm{k}_T\,q(x,\bm{k}_T ^2) = q(x)
 \end{equation}
must be given by
  \begin{equation} \label{UA}
 q(x,\bm{k}_T ^2)= \frac{1}{\pi M^2}\,\varphi _3\big( x +
\frac{\bm{k}_T ^2}{x M^2}\big)\, \theta [x(1-x)\,M^2 -\bm{k}_T ^2
]\>.
  \end{equation}
Thus the $x$ and $\bm{k}_T ^2$ dependence are intimately linked.
Moreover, the maximum size of $|\bm{k}_T|$ is bounded and
$x$-dependent, and in general $\bm{k}^2_T \,\leq\, M^2/4 \,\simeq\,
0.25$ GeV$^2$.

 Most importantly, if the functional form of $q(x_{\rm Bj})$ is known
(for consistency it must, as usual, satisfy $q(1)=0$) then from
Eq.~(\ref{SA})
   \begin{equation} \label{VA}
   \varphi _3 (\eta) = -\frac{\rm d}{{\rm d} \eta}\,\Big[\frac{q(\eta)}{\eta}\Big]\>.
  \end{equation}
Therefore we have the remarkable result that, in the $g=0$ case, the
$\bm{k}_T ^2$ dependence of $q(x,\bm{k}_T^2)$ is completely
determined by the $x_{\rm Bj}$ dependence of the standard collinear
quark density $q(x_{\rm Bj})$:
  \begin{equation} \label{VAA}
 q(x,\bm{k}_T ^2)= - \frac{1}{\pi M^2}\, \frac{\rm d}{{\rm
d}x}\,\Big[\frac{q(x)}{x}\Big]_{x=\eta}\, \theta [x(1-x)\,M^2
-\bm{k}_T ^2 ]\>, \end{equation}
where, we remind the reader,
  \begin{equation} \label{ETA}
  \eta = x + \frac{\bm{k}_T^2}{x M^2}\>.
  \end{equation}
Notice that Eqs.~(\ref{VA})-(\ref{ETA}) are in agreement with Eqs.~(47),~(48) of Ref.~\cite{Zavada:2009sk}.

As a consequence, the average transverse momentum squared,
$\langle\,\bm{k}_T^2(x)\,\rangle_q$, can be easily obtained:
\begin{equation}
 \langle\,\bm{k}_T^2(x)\,\rangle_q \>=\>\frac{M^2x^2}{q(x)}\,\int_x^1{\rm d}y\,\frac{q(y)}{y}\>.
 \label{ave-k2}
\end{equation}

For the case of polarized DIS the result is slightly more
complicated. Given the functional form of $\Delta q(x_{\rm Bj})$ we
can obtain $\varphi _8(\eta)$ via Eq.~(\ref{LA}), in agreement with Eq.~(42) of Ref.~\cite{Zavada:2007ww}:
  \begin{equation} \label{WA}
  \varphi_8 (\eta )= -\frac{1}{\eta ^3}\, \Big\{\, 3 \, \Delta q(\eta)
 -\eta \, \frac{\rm d}{{\rm d}\eta}\,\Delta q(\eta) + 2\,\int _\eta^1
\, \frac{{\rm d}\eta'}{\eta'} \,\Delta q(\eta')\Big\}\>.
   \end{equation}
The transverse momentum dependent polarized parton density is then
given, via Eq.~(\ref{GA}) and the relation between $\eta$ and
$\bm{k}_T ^2$ in Eq.~(\ref{ETA}), by
   \begin{equation} \label{WA2}
   \Delta q(x, \bm{k}_T ^2 )=  \frac{1}{\pi M^2}\,\Big[\,\frac{\bm{k}_T ^2}{x M^2} - x\, \Big]
    \,\varphi _8\big( x + \frac{\bm{k}_T ^2}{x M^2}\big)\, \theta [x(1-x)\,M^2 -\bm{k}_T ^2 ]\>,
   \end{equation}
with
   \begin{equation} \label{XA}
  \int {\rm d}^2\bm{k}_T\,\Delta q(x,\bm{k}_T ^2) =\Delta q(x)\>.
 \end{equation}
Just as for the unpolarized density, we see, via Eq.~(\ref{WA}),
that the $ \bm{k}^2_T$  dependence of $\Delta q(x,\bm{k}^2_T)$ is
completely determined by the $x_{\rm Bj}$ dependence
of the collinear polarized quark density $\Delta q(x_{\rm Bj})$.

Of great interest at present is the transversely polarized quark
density~\cite{Ralston:1979ys} (also referred to as the transversity,
see e.g.~Ref.~\cite{Barone:2001sp} for a review) which is concerned with quarks
transversely polarized along the $y$-direction inside a nucleon
transversely polarized along $OY$:
  \begin{eqnarray} \label{DB}
   \Delta _T q (x,\,\bm{k}_T) &\equiv&  q_{s_y / S_Y} (x,\,\bm{k}_T) -
    q_{-s_y/ S_Y}(x,\,\bm{k}_T) \nonumber \\
   &=&    \,\frac{\eta}{\pi M^2}\,
    \,\varphi _{11}\big( x + \frac{\bm{k}_T ^2}{x M^2}\big)\, \theta [x(1-x)\,M^2 -\bm{k}_T ^2 ]\, \cos\phi.
    \end{eqnarray}
 Here we specify the parton distributions using the physically
motivated probabilistic notation of Ref.~\cite{Anselmino:2005sh},
referring to the polarization of a parton, whose spin direction (in
its helicity frame whose axes are labelled $ox$, $oy$,
$oz$~\cite{Leader:2001gr}) is indicated by $s= s_x, s_y$ for
transverse polarization, inside a nucleon moving along $OZ$ in some
fixed frame with axes labelled $OX$, $OY$, $OZ$ (it could be the laboratory frame
or the c.m.~frame of the $\gamma^*$-nucleon collision), whose spin
direction is indicated by $S=S_X, S_Y$ for transverse polarization,
namely
 \begin{equation} \label{AB}
  \Delta q_{s /S} (x,\,\bm{k}_T)  \equiv  q_{s/S}(x,\,\bm{k}_T) - q_{-s /S} (x,\,\bm{k}_T)\>.
  \end{equation}
The connection with the more formal notation of Ref.~\cite{Mulders:1995dh}
will be given in section~\ref{tmd-others} (see also appendix C of Ref.~\cite{Anselmino:2005sh}).
In Eq.~(\ref{DB}) $\phi$ is the azimuthal angle of  $ \bm{k}_T$ in the fixed frame.

Note that
\begin{equation}
 \int {\rm d}^2\bm{k}_T \,\Delta _T q(x, \bm{k}_T) =0\>.
\label{DB2}
\end{equation}
This is because $\Delta _T q(x, \bm{k}_T)$ corresponds to having
the quark polarized along $oy$ perpendicular to \textit{its} momentum and not
along $OY$, perpendicular to the nucleon's momentum.
The $\bm{k}_T$-dependent density which does integrate to the usual
collinear transversely polarized or transversity density is one in
which the quark is polarized along the  direction $OY$ in the fixed
frame:
\begin{eqnarray}
 \Delta q _{s_Y/S_Y} (x, \bm{k}^2_T )&=& \Delta'_T q(x,
\bm{k}^2_T) \qquad \textrm{in the notation of Ref.~\cite{Barone:2001sp}} \nonumber \\
&=& h_1(x, \bm{k}^2_T )\quad\qquad  \textrm{in the notation of Ref.~\cite{Mulders:1995dh}}
\end{eqnarray}
 so that
\begin{equation} \label{GB1}
 \Delta _T q (x) = \int {\rm d}^2\bm{k}_T\,h_1(x,\bm{k}_T ^2) \>.
\end{equation}

One finds
\begin{equation} \label{GB}
\Delta'_T q(x,
\bm{k}^2_T)\equiv h_1(x, \,\bm{k}_T ^2) = \frac{x}{\pi  \, M^2}\,
    \,\varphi _{11}\big( x + \frac{\bm{k}_T ^2}{x M^2}\big)\, \theta [x(1-x)\,M^2 -\bm{k}_T ^2 ]\>.
\end{equation}
It follows that
\begin{equation} \label{Q1}
\Delta _T q(x, \bm{k}_T)=\frac{\eta}{x}\,h_1(x,\bm{k}_T ^2)\,\cos\phi\>.
\end{equation}
Moreover, from Eq.~(\ref{GB1}) one can show that
\begin{equation} \label{Q2}
\varphi _{11}(\eta) = \frac{1}{\eta^2}\, \Big\{\,\frac{2}{\eta}\Delta _T q (\eta) -
 \frac{\rm d}{{\rm d}\eta}\Delta _T q (\eta) \Big\} =
  -\,\frac{\rm d}{{\rm d}\eta}\Big[\frac{\Delta_T q(\eta)}{\eta^2}\Big]\>.
\end{equation}
Thus, via Eqs.~(\ref{GB}),~(\ref{Q1}), the $\bm{k}_T$ dependence of
both $h_1(x, \,\bm{k}_T ^2)$ and $\Delta _T q(x, \bm{k}_T)$ is
completely determined by the functional form of the standard
collinear transversity density $\Delta _T q(x)$.

 It is an intriguing question whether, in the real case of $g\neq 0$,
it is possible to have a factorized form $F(x)G(\bm{k}_T^2)$ for the
parton densities. Aside from the more complex tensorial structure
when $g\neq 0$, the functional dependence will be controlled by
expressions of the form
\begin{equation*}
\int {\rm d} k^2\> \Phi\left(\,k^2,x+\frac{\bm{k}_T^2+k^2}{xM^2}\,\right)\>,
\label{functional}
\end{equation*}
where the range of $k^2$ is limited to a small region around $k^2=0$.
While it seems unlikely that this could lead to a factorized form,
we have been unable to demonstrate this mathematically.
 \subsection{\label{tmd-others} Other transverse momentum dependent quark densities}
In this subsection we shall give expressions for several other TMD
quark densities, which, while playing no role in the
$\bm{k}_T$-integrated hadronic correlator for DIS, are important in
other processes, like semi-inclusive DIS or Drell-Yan dilepton
production, and have been widely used in the
literature~\cite{D'Alesio:2007jt,Barone:2001sp}.

Related to $A_8$ we consider quarks polarized longitudinally (i.e.
with helicity $\pm 1/2$) inside a transversely polarized nucleon
(here it is irrelevant which transverse direction is involved).
Then (see also Ref.~\cite{Mulders:1995dh})
 \begin{equation} \label{BB}
  \Delta q_{s_z /\bm{S}_T }(x,\,\bm{k}_T) \equiv  q_{+ /\bm{S}_T }(x,\,\bm{k}_T)
   - q_{- /\bm{S}_T }(x,\,\bm{k}_T)
  =  \frac{ \bm{k}_T \cdot \bm{S}_T  }{M} \, g_{1T}\>,
 \end{equation}
with
 \begin{equation} \label{CB}
 g_{1T}(x, \,\bm{k}_T ^2) = - \,\frac{2}{\pi M^2}\,
    \,\varphi _8\big( x + \frac{\bm{k}_T ^2}{x M^2}\big)\, \theta [x(1-x)\,M^2 -\bm{k}_T ^2 ]\>.
  \end{equation}

For the quark polarized in the $x$-direction, we have
 \begin{eqnarray} \label{EB}
   \Delta q_{s_x/S_Y}(x,\,\bm{k}_T) &\equiv&  q_{s_x / S_Y} (x,\,\bm{k}_T) -
    q_{-s_x/ S_Y}(x,\,\bm{k}_T) \nonumber \\
&=& \Big\{\, h_1(x, \,\bm{k}_T ^2) + \frac{\bm{k}_T ^2
}{2M^2}h_{1T}^\perp (x, \,\bm{k}_T ^2)\,\Big\} \, \sin\phi\>,
\nonumber \\
&=& - \,\frac{1}{\pi M^2}\,\Big[\,\frac{\bm{k}_T ^2}{x M^2}-x\,\Big]
\,\varphi _{11}\big( x + \frac{\bm{k}_T ^2}{x M^2}\big)\, \theta
[x(1-x)\,M^2 -\bm{k}_T ^2 ]\,\sin\phi \>.
\end{eqnarray}
One also finds that
\begin{equation} \label{FB}
h_{1T}^\perp (x,\bm{k}_T ^2) = - \,\frac{2}{\pi x \, M^2}\,
    \,\varphi _{11}\big( x + \frac{\bm{k}_T ^2}{x M^2}\big)\, \theta [x(1-x)\,M^2 -\bm{k}_T ^2 ]\>.
\end{equation}

{}For the function $h_{1T}(x,\bm{k}_T ^2)$ defined via
\begin{equation} \label{HB}
h_1 (x,\bm{k}_T ^2) \equiv  h_{1T}(x,\bm{k}_T ^2) + \frac{\bm{k}_T
^2}{2 \, M^2}\,h_{1T}^\perp (x,\bm{k}_T)\>, \end{equation}
we find
\begin{equation} \label{IB}
h_{1T}(x,\bm{k}_T ^2) = \frac{1}{\pi  \, M^2}\,\Big[\,x +
\frac{\bm{k}_T ^2}{x M^2}\,\Big]  \,\varphi _{11}\big( x +
\frac{\bm{k}_T ^2}{x M^2}\big)\, \theta [x(1-x)\,M^2 -\bm{k}_T ^2
]\>.
\end{equation}

{}Finally, for a quark polarized along the $x$-direction inside a
longitudinally polarized nucleon,
\begin{equation} \label{JB}
\Delta q_{s_x/S_Z}(x,\,\bm{k}_T ^2 ) \equiv  q_{s_x/ +}
(x,\,\bm{k}_T ^2 ) - q_{-s_x/ +}(x,\,\bm{k}_T ^2) = \frac{|\bm{k}_T|}{M}\,
h_{1L}^\perp (x,\bm{k}_T ^2)\>,
\end{equation}
with
\begin{equation} \label{KB}
h_{1L}^\perp (x,\bm{k}_T ^2) = - \,\frac{2}{\pi  \, M^2}\, \,\varphi
_{11}\big( x + \frac{\bm{k}_T ^2}{x M^2}\big)\, \theta [x(1-x)\,M^2
-\bm{k}_T ^2 ]\>.
\end{equation}

Note that for consistency in the parton model, with $g=0$, we have, in agreement with the results of EZTS,
\begin{equation} \label{LB}
\frac{x}{\eta}\,h_{1T}(x,\bm{k}_T ^2) = -
\frac{x}{2}\,h_{1L}^\perp(x,\bm{k}_T ^2) =
-\frac{x^2}{2}\,h_{1T}^\perp (x,\bm{k}_T ^2) = h_1 (x,\bm{k}_T
^2)\>.
\end{equation}
\section{\label{bounds} Bounds on the collinear densities}
In attempting to extract information on the collinear polarized
density $\Delta q (x )$ and the transversity density $\Delta _T
q(x)$ by fitting data using parameterized forms for the densities,
it is usual to impose the positivity bound
\begin{equation} \label{OBA}
|\Delta q (x )| \leq q (x)
\end{equation}
on $\Delta q (x )$, and the Soffer bound~\cite{Soffer:1994ww}
\begin{equation} \label{MBA}
|\Delta_T q (x )| \leq \frac{1}{2}\, [q (x) + \Delta q (x)]
\end{equation}
on  $\Delta _T q (x )$.
However, it turns out that while these bounds are \textit{necessary} they are not \textit{sufficient}. \\
\subsection{\label{bounds-pos} The new positivity bound}
The reason why Eq.~(\ref{OBA}) is not sufficient is that from a
probabilistic point of view one should expect
\begin{equation} \label{OB}
|\Delta q (x, \bm{k}^2_T )| \leq q (x, \bm{k}^2_T )\>,
\end{equation}
and while Eq.~(\ref{OB}) implies Eq.~(\ref{OBA}), the reverse is not
true. Actually, as we shall now show, it turns out that it is
sufficient to require just
\begin{equation} \label{OBB}
|\Delta q (x, 0 )| \leq q (x, 0)\>.
\end{equation}
For then from Eq.~(\ref{WA2})
 \begin{equation} \label{PB}
   \Delta q(x,0) = - \frac{x}{\pi M^2}\,
    \,\varphi _8 ( x ) \theta [x(1-x)\,M^2 ]\>,
   \end{equation}
and from Eq.~(\ref{UA})
\begin{equation} \label{QB}
  q(x,0)= \frac{1}{\pi M^2}\,\varphi _3( x ) \theta [x(1-x)\,M^2]\>.
  \end{equation}
Since Eq.~(\ref{OBB}) must hold for all $x$ we find that (of course $\varphi _3 (\eta )\geq 0 $)
  \begin{equation} \label{PB2}
  | \varphi _8 ( \eta )| \leq \frac{\varphi _3 ( \eta )}{\eta}\>,
  \end{equation}
  and this is sufficient to guarantee that Eq.~(\ref{OB}) holds.
Thus, we suggest that in parameterizing the collinear polarized
density $\Delta q (x )$ one should perhaps impose the stronger bound
\begin{equation}\label{NUPOS}
 \Big|\, 3 \, \Delta q(x)
 -x \, \frac{\rm d}{{\rm d}x}\,\Delta q(x) + 2\,\int _x^1 \,
\frac{{\rm d}x'}{x'} \,\Delta q(x')\,\Big| \leq
 q(x)-x\,\frac{\rm d}{{\rm d}x}\,q(x) \>,
\end{equation}
which follows from Eqs.~(\ref{PB2}),~(\ref{WA}) and (\ref{VA}) and is in agreement with Eq.~(48) of Ref.~\cite{Zavada:2007ww};
of course, one must have $q(x)-x\,q^\prime (x) \geq 0$\,.

Notice that there is a further bound,  originally derived in Ref.~\cite{Zavada:2007ww}, Eq.~(60),
\begin{equation}
\Big|\,\Delta q(x) + 2\int_x^1 \,
\frac{{\rm d}x'}{x'} \,\Delta q(x')\,\Big| \leq
 q(x)\>,
 \label{nupos-2}
\end{equation}
which can be easily obtained from our Eqs.~(\ref{KA}),~(\ref{FA}),~(\ref{SA}) and (\ref{PB2}).
\subsection{\label{bounds-soff} Reinterpretation of the Soffer bound}
Soffer derived his bound on the collinear transversity density~\cite{Soffer:1994ww}
\begin{equation} \label{MB}
|\Delta _T q (x )| \leq \frac{1}{2}\, [q (x ) + \Delta q (x )]\>,
\end{equation}
by noticing  the analogy between the quark correlator diagram and
the diagram for the $u$-channel absorptive part of \textit{forward}
elastic quark-nucleon scattering. But collinear parton densities
include all partons with transverse momentum $\bm{k}^2_T\leq Q^2 $
and thus do not consist of strictly forward moving partons. Hence,
in the light of the existence of intrinsic $\bm{k}_T $,  Soffer's
analogy  could be said to have been misinterpreted, and the correct
bound should read
\begin{equation} \label{NB}
|\Delta _T q (x, 0 )|\>\equiv\> |\Delta '_T q (x, 0 )| \>\leq\>
\frac{1}{2}\, [q (x, 0 ) + \Delta q (x,0 )]\>.
\end{equation}
Now, from Eqs.~(\ref{Q1}),~(\ref{GB}) we have
  \begin{equation} \label{QB2}
   \Delta _T q (x,0 ) \>\equiv\> \Delta' _T q (x,0 ) \>=\>  \frac{x}{\pi M^2}\,
    \,\varphi _{11}( x )\, \theta [x(1-x)\,M^2]\>,
    \end{equation}
and substituting this and Eqs.~(\ref{PB}),~(\ref{QB})  into Eq.~(\ref{NB}) we obtain
  \begin{equation} \label{RB}
 |\varphi _{11}( \eta )| \leq \frac{1}{2 \eta}\, [\varphi _{3}( \eta)
 - \eta \, \varphi _{8}( \eta )]\>.
 \end{equation}

 We suggest that in parameterizing models for the collinear
transversity density $\Delta_T q(x)$, the scalar functions $\varphi
_{3,8,11}$ should be calculated in terms  of $q(x)$, $\Delta q(x)$ and
$\Delta_T q(x)$, via Eqs.~(\ref{VA}),~(\ref{WA}),~(\ref{Q2}), and,
perhaps, should be forced to satisfy the bound in Eq.~(\ref{RB}).

{}Finally, we have also explicitly verified that the bounds in
Eqs.~(\ref{PB2}), (\ref{RB}) are equivalent to the bounds on the
$\bm{k}_T$-dependent functions derived in Ref.~\cite{Bacchetta:1999kz},
in the $g=0$ limit.
\subsection{\label{bounds-pdf} Collinear and TMD quark distributions}
As a simple and illustrative example of our results, we
consider here some of the consequences of imposing Lorentz
invariance on the $\bm{k}_T$-dependence of TMD parton distributions.
We will start from a well-known set of unpolarized and
longitudinally polarized distributions extracted within the standard
collinear approach,  namely the GRV98~\cite{Gluck:1998xa} and
GRSV2000~\cite{Gluck:2000dy} sets respectively, and show the crucial
role played by the new bounds. Since we are working in the $g=0$
approximation, we will consistently consider only $u$ and $d$ quark
distributions, neglecting sea quarks. Moreover, since we are here
mainly interested in the intrinsic $\bm{k}_T$ dependence, we choose
as reference scale a relatively low one, $Q^2=2$ GeV$^2$, in order
to exclude large effects due to perturbative evolution. From these
parameterizations for $q(x,Q^2)$ and $\Delta q(x,Q^2)$ we therefore
generate, via Eq.~(\ref{VAA}) and Eqs.~(\ref{WA}),~(\ref{WA2}) the
$\bm{k}_T$-dependent distributions $q(x,\bm{k}_T^2,Q^2)$ and $\Delta
q(x,\bm{k}_T^2,Q^2)$.
\begin{figure*}[b]
 \includegraphics[angle=-90,width=0.7\textwidth]{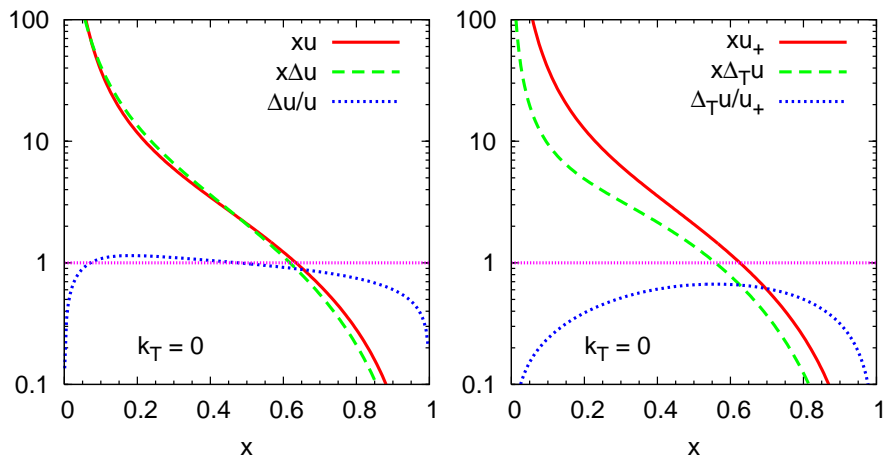}
 \caption{ Left panel: unpolarized (solid line), longitudinally polarized
(dashed line) distributions and their ratio (dotted line) for up quarks
as a function of $x$ at $|\bm{k}_T| = 0$ and $Q^2 = 2$ GeV$^2$,
obtained via Eq.~(\ref{VAA}) and Eqs.~(\ref{WA}),~(\ref{WA2})
starting from $u(x)$ [GRV98] and $\Delta u(x)$ [GRSV2000]. Note the
violation of the positivity bound, Eq.~(\ref{OBB}). Right panel: Soffer-type
bound, $u_+ \equiv (u + \Delta u)/2$, (solid line), transversity
distribution (dashed line) and their ratio (dotted line)
for up quarks  as a function of $x$ at $|\bm{k}_T| = 0$ and $Q^2 = 2$ GeV$^2$,
obtained via Eqs.~(\ref{Q2}),~(\ref{QB2}) starting from
$\Delta_Tu(x)$ of Ref.~\cite{Anselmino:2008jk}.
 \label{u-k0-x} }
 \end{figure*}
 \begin{figure*}[t]
 \includegraphics[angle=-90,width=0.7\textwidth]{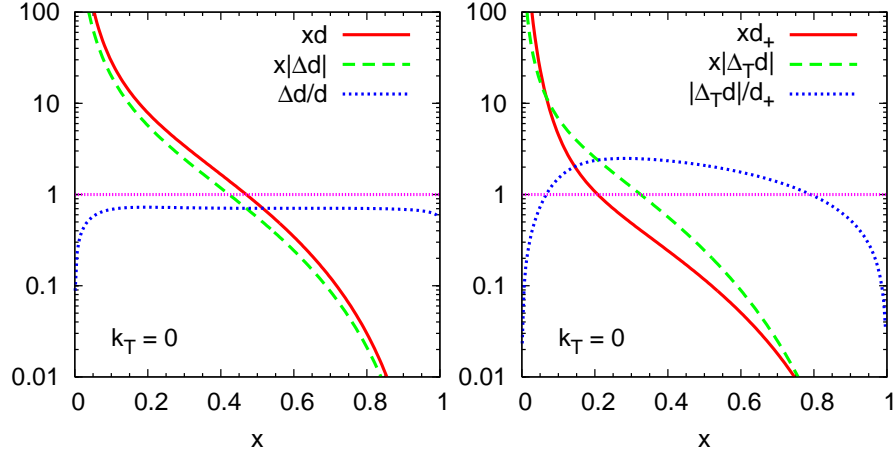}
 \caption{Same as in Fig.~\ref{u-k0-x}, but for down quarks. Note, in the right panel,
  the strong violation of the Soffer-type bound, Eq.~(\ref{NB}).
 \label{d-k0-x} }
 \end{figure*}

Concerning transversity, we have to recall that at present no
parameterization in the standard collinear approach is available.
Double transverse spin asymmetries, $A_{TT}$, in the Drell-Yan
processes $p^\uparrow p^\uparrow\to \ell^+\ell^- + X$, or
$p^\uparrow \bar{p}^\uparrow \to \ell^+\ell^- + X$, are considered
the best tool to this end, but are unfortunately still out of reach
for present experimental setups. In the meantime, the study of
transverse single spin asymmetries in polarized SIDIS processes in
the TMD approach has been found to be the most promising way to get
information on the unintegrated, $\bm{k}_T$-dependent transversity
distribution~\cite{Anselmino:2007fs}. We will therefore utilize in
our discussion the first ever parameterization available for the
transversity of $u$ and $d$ quarks obtained by fitting, in the TMD
approach, the so-called Collins azimuthal asymmetries measured in
SIDIS pion and kaon production by the HERMES and COMPASS
experiments. We have to stress, however, that this parameterization
has been obtained by assuming a factorized $x$ and
$\bm{k}_T$-dependence for the TMD distributions, which is at
variance with our findings in the simplified, $g=0$, limit.
Given the still poor knowledge of the transversity distribution, we
believe however that, despite this inconsistency, it is of interest
to investigate the consequences of our approach. Starting from the
most updated parameterization of the transversity distribution for
valence quarks, Ref.~\cite{Anselmino:2008jk}, and using
Eqs.~(\ref{Q2}),~(\ref{GB}), we therefore generate the $g=0$,
$\bm{k}_T$-dependent distribution $\Delta'_T q(x,\bm{k}_T^2,Q^2)$.

\begin{figure*}[b]
 \includegraphics[angle=-90,width=0.7\textwidth]{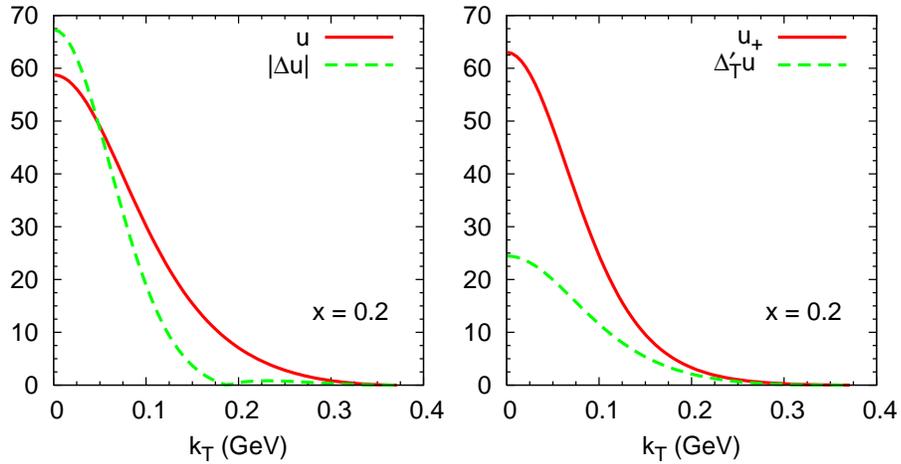}
 \caption{ Left panel: unpolarized (solid line) and longitudinally polarized
(dashed line) distributions for up quarks as a function of $|\bm{k}_T|$ at
$x = 0.2$ and $Q^2 = 2$ GeV$^2$, obtained via Eq.~(\ref{VAA}) and
Eqs.~(\ref{WA}),~(\ref{WA2}) starting from $u(x)$ [GRV98] and
$\Delta u(x)$ [GRSV2000]. Right panel: Soffer-type bound (solid line) and
transversity distribution, $\Delta'_T u$, (dashed line) for up quarks,
as a function of $|\bm{k}_T|$ at $x = 0.2$ and $Q^2 = 2$ GeV$^2$,
obtained via Eqs.~(\ref{GB}),~(\ref{Q2}) starting from
$\Delta_Tu(x)$ of Ref.~\cite{Anselmino:2008jk}.
 \label{u-x02-k} }
\end{figure*}

\begin{figure*}[t]
 \includegraphics[angle=-90,width=0.7\textwidth]{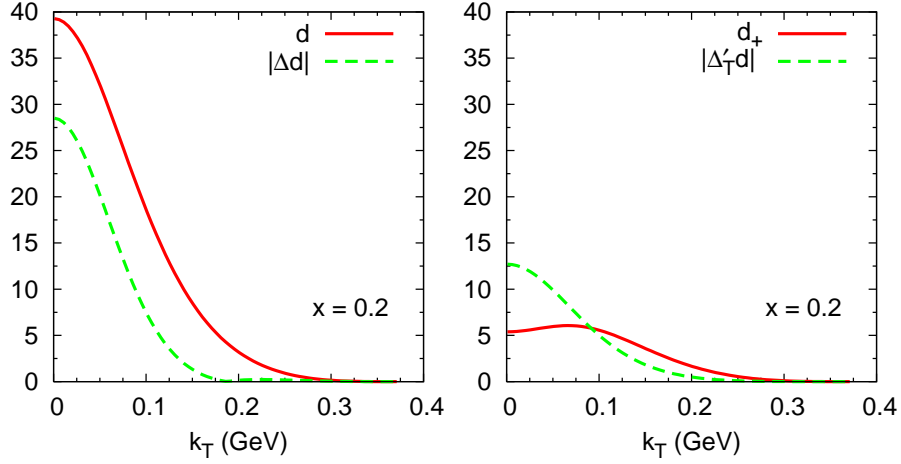}
 \caption{ Same as in Fig.~\ref{u-x02-k}, but for down quarks.
 \label{d-x02-k} }
\end{figure*}

\begin{figure*}[b]
 \includegraphics[angle=-90,width=0.5\textwidth]{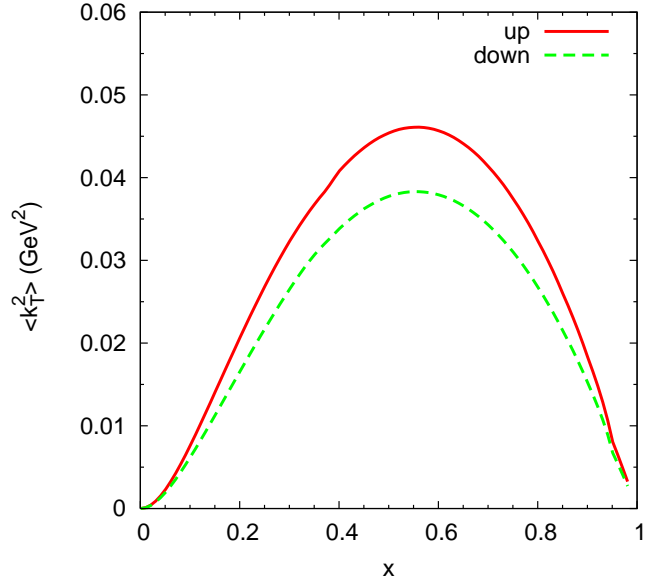}
 \caption{ Mean transverse momentum squared, $\langle\,\bm{k}_T^2(x)\,\rangle_q$,
Eq.~(\ref{ave-k2}), as a function of $x$, for up (solid line) and
down (dashed line) quarks at $Q^2=2$ GeV$^2$.
 \label{kt2-x} }
\end{figure*}

Our results are summarized in Figs.~\ref{u-k0-x}-\ref{kt2-x}. In
Fig.~\ref{u-k0-x}, left panel, we show the unpolarized and
longitudinally polarized $u$ quark distributions, and their ratio,
as a function of $x$ at fixed $|\bm{k}_T|=0$. Although the collinear
parton distributions adopted here fulfill the usual positivity
bound, Eq.~(\ref{OBA}), we see that the stronger positivity bound of
Eq.~(\ref{OBB}) is indeed violated for some $x$ values. This is most
clearly seen by looking at the ratio $\Delta u/u$ (dotted line),
where the violation reachs about 15\%. In Fig.~\ref{u-k0-x}, right
panel, we also show, again for $u$ quarks, the transversity
distribution $\Delta_T u$ and the positive-helicity distribution
$u_+=(u+\Delta u)/2$ and their ratio (related to the Soffer bound),
as a function of $x$ at fixed $|\bm{k}_T|=0$. In this case also the
stronger, Soffer-type, bound, Eq.~(\ref{NB}), is fulfilled.

Similar results are shown in Fig.~\ref{d-k0-x} for $d$ quarks. This
time, while the stronger positivity bound derived in this section is
always fulfilled, the Soffer-type bound, Eq.~(\ref{NB}), is instead
violated over a large range of $x$ values.

In Fig.~\ref{u-x02-k} we show, again for $\Delta u(x,\bm{k}_T^2)$ in
the left panel and for $\Delta'_T u(x,\bm{k}_T^2)$ in the right
panel, the $|\bm{k}_T|$ dependence of the $u$-quark distributions at
a fixed value of $x$ in the valence region, $x=0.2$. For the
longitudinal distribution, we see that the violation of the stronger
positivity bound at $|\bm{k}_T|=0$ persists up to some relatively
large values of $|\bm{k}_T|$. A similar behaviour should result for
any value of $x$ in the region of violation of the positivity bound
at $|\bm{k}_T|=0$, shown in the left panel of Fig.~\ref{u-k0-x}.
Notice also the node in $\Delta u$, which is a general feature of
$\Delta q$ coming from Eq.~(\ref{WA2}). Finally, notice from the
right panel that, as expected, the validity of the Soffer-type bound at
$|\bm{k}_T|=0$ guarantees its validity at any allowed value of
$|\bm{k}_T|$. Analogous results for $d$ quarks are presented in
Fig.~\ref{d-x02-k}, and similar comments hold on the violation, this
time, of the Soffer-type bound.

For completeness, in Fig.~\ref{kt2-x} we show the resulting mean transverse
momentum squared, $\langle\,\bm{k}_T^2(x)\,\rangle_q$,
Eq.~(\ref{ave-k2}), as a function of $x$, for $u$ and $d$ quarks, at
the same reference scale, $Q^2=2$ GeV$^2$. We briefly note that the
shape of $\langle\,\bm{k}_T^2(x)\,\rangle_q$ is quite reasonable and
approximately flavour-independent, although its size is considerably
smaller than expected from available phenomenological analyses
within the TMD approach~\cite{D'Alesio:2004up,Anselmino:2005nn}.

This is perhaps a hint that the larger values of $\langle\,\bm{k}_T^2\,\rangle$ needed in the
phenomenological analyses may be due to hard gluon emission and may
therefore not be a measure of the size of the genuinely intrinsic
transverse momentum. It would be interesting to construct models
based upon such a picture.
\section{\label{concl} Conclusions}
We have shown that the simple parton model has a very rich structure when due care
is given to its Lorentz properties.
On the basis that \emph{kinematic} relations should not depend on the value of the
strong coupling $g$, and thus should be valid when $g=0$,
we have been able to give a simple derivation of the exact target mass corrections
for polarized DIS, without recourse to the operator product expansion,
and have verified that the Wandzura-Wilczek relation holds when target mass
corrections are included. We have also demonstrated that
that there is a surprising and  intimate connection between the intrinsic $\bm{k}_T$
dependence of parton densities and the functional form of the
$\bm{k}_T$-integrated collinear densities. Indeed the $\bm{k}_T$ dependence can actually be
\emph{derived} from the collinear densities. Using some
published versions of the collinear densities we have studied the  implications of
this relationship and found that our $\bm{k}_T$-dependent distributions,
thus derived, are much narrower than those used in phenomenological work i.e.~our
$\langle\,\bm{k}_T^2\,\rangle$ is much smaller. This suggests that
some of the $\bm{k}_T$ dependence usually described as intrinsic, may, in fact, be due to
gluon radiation.

One outcome of our work is the clear demonstration that the factorized form
$g(x)f(\bm{k}_T^2)$ often used in phenomenological studies,
is untenable, \emph{at least when $g=0$}. While not proved, it seems likely that this
result is quite general, though, as commented on, a factorized form does seem a reasonable starting
point for the analysis of the presently available data.
In addition, because of this interrelationship between $\bm{k}_T$ and $x$ dependence, we
have been able to derive new positivity and Soffer-type
bounds on the collinear densities, which are stronger than the those usually
utilized in DIS  and SIDIS analyses. We have shown numerically that some of the
 published collinear parton densities, which satisfy the traditional bounds, do not
always satisfy these stronger bounds. The particular case of the down-quark transversity
may be of some interest.
\begin{acknowledgments}
E.L.~is grateful to Dr.~Ekaterina Christova for many discussions in
the early stages of this work, and to the Department of Physics of the University of Cagliari for the
hospitality extended to him during several visits.
\end{acknowledgments}


\begin{thebibliography}{10}
%
\bibitem{Wandzura:1977qf}
S.~Wandzura and F.~Wilczek, Phys. Lett. {\bf B72}, 195 (1977).

\bibitem{Nachtmann:1973mr}
O.~Nachtmann, Nucl. Phys. {\bf B63}, 237 (1973).

\bibitem{Georgi:1976ve}
H.~Georgi and H.~D. Politzer, Phys. Rev. {\bf D14}, 1829 (1976).

\bibitem{Matsuda:1979ad}
S.~Matsuda and T.~Uematsu, Nucl. Phys. {\bf B168}, 181 (1980).

\bibitem{Wandzura:1977ce}
S.~Wandzura, Nucl. Phys. {\bf B122}, 412 (1977).

\bibitem{Piccione:1997zh}
A.~Piccione and G.~Ridolfi, Nucl. Phys. {\bf B513}, 301 (1998), {{\tt hep-ph/9707478}}.

\bibitem{Blumlein:1998nv}
J.~Bl\"{u}mlein and A.~Tkabladze, Nucl. Phys. {\bf B553}, 427 (1999), {{\tt hep-ph/9812478}}.

\bibitem{Ellis:1982cd}
R.~K. Ellis, W.~Furmanski, and R.~Petronzio, Nucl. Phys. {\bf B212}, 29 (1983).

\bibitem{D'Alesio:2007jt}
U.~D'Alesio and F.~Murgia, Prog. Part. Nucl. Phys. {\bf 61}, 394 (2008), {{\tt 0712.4328[hep-ph]}}.

\bibitem{Soffer:1994ww}
J.~Soffer, Phys. Rev. Lett. {\bf 74}, 1292 (1995), {{\tt hep-ph/9409254}}.

\bibitem{Landshoff:1970ff}
P.~V. Landshoff, J.~C. Polkinghorne, and R.~D. Short, Nucl. Phys. {\bf B28}, 225 (1971).

\bibitem{Franklin:1976dm}
J.~Franklin, Phys. Rev. {\bf D16}, 21 (1977).

\bibitem{Jackson:1989ph}
J.~D. Jackson, G.~G. Ross, and R.~G. Roberts, Phys. Lett. {\bf B226}, 159 (1989).

\bibitem{Zavada:1996kp}
P.~Zavada, Phys. Rev. {\bf D55}, 4290 (1997), {{\tt hep-ph/9609372}}.

\bibitem{Zavada:2002uz}
P.~Zavada, Phys. Rev. {\bf D67}, 014019 (2003), {{\tt hep-ph/0210141}}.

\bibitem{Zavada:2007ww}
P.~Zavada, Eur. Phys. J. {\bf C52}, 121 (2007), {{\tt 0706.2988[hep-ph]}}.

\bibitem{Efremov:2004tz}
A.~V. Efremov, O.~V. Teryaev, and P.~Zavada, Phys. Rev. {\bf D70}, 054018 (2004), {{\tt hep-ph/0405225}}.

\bibitem{Efremov:2009ze}
A.~V. Efremov, P.~Schweitzer, O.~V. Teryaev, and P.~Zavada, 
Phys. Rev. {\bf D80}, 014021 (2009), {{\tt 0903.3490[hep-ph]}}.

\bibitem{Zavada:2009sk}
P.~Zavada, {{\tt 0908.2316[hep-ph]}}.

\bibitem{Ralston:1979ys}
J.~P. Ralston and D.~E. Soper, Nucl. Phys. {\bf B152}, 109 (1979).

\bibitem{Leader:1979ys}
E.~Leader, private communication to Ralston and Soper (1979).

\bibitem{Mulders:1995dh}
P.~J. Mulders and R.~D. Tangerman, Nucl. Phys. {\bf B461}, 197 (1996), {{\tt hep-ph/9510301}}.

\bibitem{Anselmino:1994gn}
M.~Anselmino, A.~Efremov, and E.~Leader, Phys. Rep. {\bf 261}, 1 (1995), {{\tt hep-ph/9501369}}.

\bibitem{Accardi:2008pc}
A.~Accardi and W.~Melnitchouk, Phys. Lett. {\bf B670}, 114 (2008), {{\tt 0808.2397[hep-ph]}}.

\bibitem{Anselmino:2005sh}
M.~Anselmino et al., Phys. Rev. {\bf D73}, 014020 (2006), {{\tt hep-ph/0509035}}.

\bibitem{Barone:2001sp}
V.~Barone, A.~Drago, and P.~G. Ratcliffe, Phys. Rep. {\bf 359}, 1 (2002), {{\tt hep-ph/0104283}}.

\bibitem{Leader:2001gr}
E.~Leader, Spin in Particle Physics, Cambridge University Press (2001).

\bibitem{Bacchetta:1999kz}
A.~Bacchetta, M.~Boglione, A.~Henneman, and P.~J. Mulders, 
Phys. Rev. Lett. {\bf 85}, 712 (2000), {{\tt hep-ph/9912490}}.

\bibitem{Gluck:1998xa}
M.~Gl\"{u}ck, E.~Reya, and A.~Vogt, Eur. Phys. J. {\bf C5}, 461 (1998), {{\tt hep-ph/9806404}}.

\bibitem{Gluck:2000dy}
M.~Gl\"{u}ck, E.~Reya, M.~Stratmann, and W.~Vogelsang,
Phys. Rev. {\bf D63}, 094005 (2001), {{\tt hep-ph/0011215}}.

\bibitem{Anselmino:2007fs}
M.~Anselmino et al., Phys. Rev. {\bf D75}, 054032 (2007), {{\tt hep-ph/0701006}}.

\bibitem{Anselmino:2008jk}
M.~Anselmino et al., Nucl. Phys. Proc. Suppl. {\bf 191}, 98 (2009), {{\tt 0812.4366[hep-ph]}}.

\bibitem{D'Alesio:2004up}
U.~D'Alesio and F.~Murgia, Phys. Rev. {\bf D70}, 074009 (2004), {{\tt hep-ph/0408092}}.

\bibitem{Anselmino:2005nn}
M.~Anselmino et al., Phys. Rev. {\bf D71}, 074006 (2005),  {{\tt hep-ph/0501196}}.

\end{thebibliography}
%

\end{document}